\documentclass[final,5p,times,twocolumn]{elsarticle}
\usepackage[colorinlistoftodos,textwidth=4cm,shadow]{todonotes}
\usepackage{amsmath}
\usepackage[noend]{algorithmic}
\usepackage{lipsum}

\journal{Computer Physics Communications}

\newcounter{Ivan}

\newcounter{Misha}

\begin{document}
\begin{frontmatter}

\title{Low rank approximations for the DEPOSIT computer code.}

\author[skoltech]{Mikhail S. Litsarev}
\ead{m.litsarev@skolkovotech.ru}
\author[skoltech,inm]{Ivan V. Oseledets}
\address[skoltech]{Skolkovo Institute of Science and Technology,
Novaya St. 100, Skolkovo, Odintsovsky district, 143025 
Moscow Region, Russia}
\address[inm]{Institute of Numerical Mathematics,
Gubkina St. 8, 119333 Moscow, Russia}

\begin{abstract}
We present an efficient technique based on low-rank separated
approximations 
for the computation of three-dimensional integrals
in the computer code DEPOSIT that describes ion-atomic collision
processes. 
Implementation of this technique decreases the total computational
time by a factor of $\sim 10^3$.
The general concept can be applied to more complicated models.
%Basic concepts are considered and 

\end{abstract}


\end{frontmatter}

\today

\section{Introduction.}
\label{IntroSect}
The computer code DEPOSIT~\cite{litsarev-cpc-2013} is intended to describe
ion-atomic collision processes.
It allows to calculate total and multiple electron loss
cross sections $\sigma$ and $\sigma_m$
($m$ is the number of ionized electrons),
the deposited energies $T(b)$,
($b$ is the impact parameter of the projectile ion)
and ionization probabilities $P_m(b)$.
It is based on the energy deposition model
introduced by N.~Bohr~\cite{bohr1915} and developed further by
A.~Russek and J.~Meli~\cite{RussekMeli1970}, 
C.L. Cocke~\cite{cocke_pra1979},
and V.P. Shevelko~\textit{at al.}~\cite{litsarev-jpb-2008}.
Theoretical development of the DEPOSIT is
presented in \cite{litsarev-jpb-2008, litsarev-nimb-2009,
litsarev-jpb-2009, litsarev-jpb-2010}.
Examples of calculations are reported
in~\cite{litsarev-springer-2012, litsarev-hci-2012, uspekhi2013, litsarev-jpb-2014}.
Detailed description of
the code and user guide are given in~\cite{litsarev-cpc-2013}.

The cross sections and ionization probabilities needed for 
estimation of the losses and lifetimes of fast ion beams, 
background pressures and pumping requirements
in accelerators and storage rings
are, in fact, functionals of the deposited energy~$T(b)$, 
which in turn is a three-dimensional integral over the coordinate
space. To calculate any of these parameters one has to compute
$T(b)$ in all points of the $b$-mesh.

The integral $T(b)$ is a bottleneck of the program, and 
it is required to be done as fast as possible.
In the previous work~\cite{litsarev-cpc-2013} an advanced 
quadrature technique was used, and the computational time
has appeared to be much faster in comparison 
with direct usage of uniform meshes. 
It takes several seconds to compute one point $T(b)$ for 
one atomic shell at fixed~$b$.
For complex ions, the total computation takes few 
hours on one processor core and is not enough fast. 
To overcome this issue a fully scalable parallel variant of the
algorithm was proposed. Nevertheless, the 
computational time is still large.

In this work, we present an entirely different approach for computing
$T(b)$ in many points of the $b$-mesh,  based on low rank
approximations of matrices and tensors. The main idea is to
approximate the functions to be integrated by a sum of products of
univariate functions, effectively decreasing the dimensionality of the
problem. This involves active usage of numerical and analytical
tools. 

The definition of $T(b)$ involves a function of two variables (the energy gain $\Delta E$ during an ion-atomic collision)
and a function of three variables (electron density in Slater-type approximation).
Details and definitions are given in Section~\ref{TbDefSect}.  The
integral is computed in Cartesian coordinates, which are better suited
for the construction of separable representation than spherical
coordinates used in the original DEPOSIT code.

In Section~\ref{LowRankSect}
for a function of two
variables we use the pseudo-skeleton
decomposition of matrices~\cite{tee-mosaic-1996,
gtz-psa-1997,gtz-maxvol-1997}
computed via a variant of the incomplete cross
approximation algorithm \cite{tee-cross-2000}. 
We show numerically that the function in question 
can be well-approximated by a separable
function in Section~\ref{NumericSect}. 
Thus, the approximation can be computed in $\mathcal{O}(n)$
time, where $n$ is the number of grid points in one dimension.

In Section~\ref{RhoExpSect} the Slater-type 
function of three variables is decomposed by 
the exponential sums approach~\cite{beylkin-expsum-2005,beylkin-expsumrev-2010}. 
The integral is immediately reduced to a two-dimensional one of a
simpler structure. 

Combining these two representations 
we obtain in Section~\ref{FastTb2DSect} 
an efficient algorithm with $\mathcal{O}(n)$ scaling, 
in comparison with $\mathcal{O}(n^3)$ complexity for direct integration 
over a three-dimensional mesh.
The computation  of $T(b)$ on the whole $b$-mesh takes less then one
minute and total speedup of the program is about $\sim 10^3$ times.
Illustrative examples are given in Section~\ref{NumericSect}.

\noindent
All the equations related to the physical model
are written in atomic units.

\section{Numerical procedure}
\label{sectTb}
\subsection{Statement of the problem}
\label{TbDefSect}
The deposited energy $T(b)$ is defined as a three-dimensional
integral over coordinate space centered in the projectile ion.
\begin{equation}
\label{Tb3DIntegral} 
T(b)=\sum_{\gamma} \int \Delta E_{\gamma}(p)
\, \rho_{\gamma}(r)
\, d^{3} \mathbf{r}.
\end{equation}
The sum here is over all atomic shells denoted
by $\gamma=nl$, $n$~is the
\emph{principle quantum number} and $l$ is the
\emph{orbital quantum number}.
The electron density $\rho_{\gamma}(r)$ is taken 
in a Slater-type approximation
\begin{equation}
\label{rhoDef}
\rho_{\gamma}(r)=C_{\gamma} r^{\alpha_{\gamma}} e^{-2\beta_\gamma r} 
\end{equation}
with integer $\alpha_{\gamma}$, real positive $\beta_{\gamma}$
and normalization condition
\begin{equation}
\int_{0}^{\infty} \rho_{\gamma}(r) dr=N_{\gamma},
\end{equation}
where $N_{\gamma}$ is the number of electrons 
in a $\gamma$-shell.
The gain of kinetic energy $\Delta E_{\gamma}$ is
a smooth finite function of parameter $|\mathbf{p}|$
without any singularities.
The impact parameter $\mathbf{p}$ of the ion's electron
is a function of $\mathbf{b}$ and $\mathbf{r}$. 
In frame of the moving projectile the following equality holds
\begin{equation}
\label{pbrrelation}
p^2=(b-r\cos\theta)^2 + (r\cos\varphi \sin \theta)^2.
\end{equation}
For details we refer the reader to the paper~\cite{litsarev-cpc-2013}.
In Cartesian coordinates $\Delta E_{\gamma}(p)$
as a function of parameter $p$
depends only on $x$ and $z$ as it follows from the 
equation~(\ref{pbrrelation})
\begin{equation}
\label{TbDef}
T_{\gamma}(b)=\iiint \! \Delta E_{\gamma}(x,z - b) \rho_{\gamma}(x,y,z) dx dy dz.
\end{equation}
Thus, we need to compute the integral~\eqref{TbDef}.
From here and bellow index $\gamma$ will be skipped
for the sake of simplicity and only one shell will be considered
in the following equations.

\begin{table*}[t]
\caption{
Ranks $r$ of the decomposition~\eqref{Eextand}
calculated by the incomplete cross approximation
algorithm~\cite{tee-cross-2000}
for the energy gain $\Delta E(x, \tilde z)$.
Two cases are considered:
collision of $Au^{26+}$ ions with the Oxigen atom
at a collision energy $E=6.5$ MeV/u
and collision of $U^{28+}$ ions with the Xenon atom
at a collision energy $E=2.5$ MeV/u.
Number of the $x$-mesh points is taken equal to $2N+1$,
number of the $\tilde z$-mesh points is taken equal to
$3N+1$ in correspondence to 
the equations~\eqref{xiPoints} and~\eqref{ztildakPoints}, $a_x=a_z=8$.
Accuracy $\varepsilon$ means relative error of the approximation
in the Frobenious norm.
The calculations were carried out on $1.3$ GHz Intel Core i5 processor.
Column $T_{\mbox{\footnotesize{cross}}}$
corresponds to the time the cross algorithm takes.
The numerical results confirm the almost linear scaling of the approach in $N$.
}
\label{Table1Ranks}
\centering
\begin{tabular}{cccccccccccccc}
\hline
\hline
System & $\gamma$-Shell & $r$ & $T_{\mbox{\footnotesize{cross}}}$ (sec) &
 $\varepsilon$ & $N$ & $r$ & $T_{\mbox{\footnotesize{cross}}}$ (sec) & 
 $\varepsilon$  & $N$ & $r$ & $T_{\mbox{\footnotesize{cross}}}$ (sec) & 
 $\varepsilon$  & $N$ \\
\hline
$Au^{26+}+ O$ & $4df^{17}$ & $13$ & $0.21$ & $10^{-6}$ & $1024$ 
& $21$ & $0.42$ & $10^{-9}$ & $1024$ & 24 & 2.41 & $10^{-9}$ & 4096 \\
 & $4sp^{8}$ & $13$ & $0.21$ & &  & $21$ & $0.33$ & & & 24 & 2.40 & & \\
 & $3d^{10}$ & $14$ & $0.19$ & &  & $22$ & $0.42$ & & & 26 & 2.58 & & \\
 & $3sp^{8}$ & $16$ & $0.25$ & &  & $24$ & $0.54$ & & & 29 & 2.64 & & \\
 & $2sp^{8}$ & $17$ & $0.28$ & &  & $25$ & $0.56$ & & & 30 & 2.71 & & \\
 & $1sp^{2}$ & $17$ & $0.27$ & &  & $25$ & $0.56$ & & & 30 & 2.70 & & \\
\hline
$U^{28+}+ Xe$ & $5sp^{4}$ & $14$ & $0.20$ & $10^{-6}$ & $1024$ 
& $22$ & $0.50$ & $10^{-9}$ & $1024$ & 26 & 2.17 & $10^{-9}$ & 4096 \\
 & $4df^{24}$ & $15$ & $0.23$ & & & $24$ & $0.52$ & & & 27 & 2.58 & & \\
 & $4sp^{8}$ & $17$ & $0.28$ & &  & $25$ & $0.55$ & & & 30 & 2.69 & & \\
 & $3d^{10}$ & $17$ & $0.27$ & &  & $25$ & $0.54$ & & & 30 & 2.77 & & \\
 & $3sp^{8}$ & $17$ & $0.27$ & &  & $25$ & $0.55$ & & & 30 & 2.75 & & \\
 & $2sp^{8}$ & $17$ & $0.26$ & &  & $25$ & $0.54$ & & & 30 & 2.71 & & \\
 & $1sp^{2}$ & $17$ & $0.26$ & &  & $25$ & $0.55$ & & & 30 & 2.69 & & \\
\hline
\hline
\end{tabular}
\end{table*}

\subsection{Low rank approximation.}
\label{LowRankSect}
Let $F(x, y)$ be a function of two variables $x, y$ where point $(x, y)$ is
in a certain rectangle $[a_x,b_x] \otimes [a_y,b_y]$. 
The function is said to be in the
\textit{separated form}
if it can be represented as a sum of products of univariate functions:
\begin{equation}
\label{FxyCanonical}
F(x,y)=\sum_{\alpha=1}^r \sigma_{\!\alpha} \,u_{\alpha}(x)g_{\alpha}(y).
\end{equation}
The minimal number $r$ such that \eqref{FxyCanonical} exists will be called \textit{separation rank}. 
%The coefficients $\sigma_{\alpha}$ are in general complex numbers.
Direct generalization of \eqref{FxyCanonical} to multivariate functions
is referred to as canonical polyadic (CP, also known as
CANDECOMP/PARAFAC) \cite{kolda-review-2009}. 

If the function is in the separated form, 
the integration is simplified a lot. Indeed,
\begin{equation}
\label{lrint:2dex}
\iint F(x,y) dx dy = \sum_{\alpha=1}^r \sigma_{\alpha} 
\! \int_{a_x}^{b_x} \!\!\! u_{\alpha}(x) dx \! \int_{a_y}^{b_y} \!\!\! g_{\alpha}(y) dy,
\end{equation}
and the problem is reduced to the computation of one-dimensional integrals,
which can be computed using fewer quadrature points than the original integral.

The discretization of one-dimensional integrals in \eqref{lrint:2dex}
by some quadrature formula with nodes
$x_i \in [a_x, b_x]$, 
$i = 1,\ldots, n$, $y_j \in [a_y, b_y]$, 
$j = 1,\ldots,m$
and weights $w^{(x)}_i$, $w^{(y)}_j$, 
leads to the approximation
\begin{equation}
\label{IntegralS1D1D}
\iint F(x,y) dx dy \approx \sum_{\alpha=1}^r \sigma_{\alpha}
\sum_{i=1}^n w^{(x)}_i u_{\alpha}(x_i) 
\sum_{j=1}^m w^{(y)}_j g_{\alpha}(y_j).
\end{equation}
On the other hand, direct two-dimensional
quadrature with separated weights in $x$ and $y$
can be used for the original integral: 
\begin{equation}
\label{integral2D}
\iint F(x,y) dx dy \approx 
\sum_{i=1}^n w^{(x)}_i 
\sum_{j=1}^m w^{(y)}_j 
F(x_i,y_j).
\end{equation}
Comparison of two representations
\eqref{IntegralS1D1D} and \eqref{integral2D}
leads to the following discrete approximation problem
\begin{equation}
\label{lrint:2ddiscr}
F(x_i, y_j) \approx 
\sum_{\alpha=1}^r \sigma_{\alpha} u_{\alpha}(x_i) g_{\alpha}(y_j),
\end{equation}
which is a discrete analogue of \eqref{FxyCanonical}.  Equation \eqref{lrint:2ddiscr} can be written in the matrix form:
\begin{equation}
A \approx U \Sigma G^{\top},
\end{equation}
where $A$ is an $n \times m$ matrix with elements $A_{ij}=F(x_i, y_j)$, 
$U$ is an $n \times r$ matrix with elements $U_{i\alpha}=u_{\alpha}(x_i)$, 
$G$ is an $m \times r$ matrix with elements $G_{\! j\alpha}=g_{\alpha}(y_j)$
 and $\Sigma$ is an $r \times r$ diagonal matrix with elements $\sigma_{\alpha}$ on the diagonal. This is a standard \emph{low-rank approximation problem} for a given matrix.
Provided that a good low-rank approximation exists, there are very efficient \emph{cross approximation algorithms} \cite{tee-cross-2000,bebe-2000} that need only $\mathcal{O}((n + m)r)$ elements of a matrix to be computed. 

By using of our implementation of the cross 
approximation algorithm we decompose
the energy gain $\Delta E(x,\tilde z)$ in the form~\eqref{lrint:2ddiscr}.
In Table~\ref{Table1Ranks} 
the ranks $r$ and other numerical parameters
are given for particular systems. 
Description of these parameters can 
be found in Section~\ref{NumericSect}.

\subsection{Exponential sums.}
\label{RhoExpSect}
For a function $\rho(x, y, z)$ defined in \eqref{rhoDef} 
the separation of variables
can be done analytically~\cite{beylkin-expsum-2005,
 hackbra-expsum-2005, GHK-ten_inverse_ellipt-2005,beylkin-expsumrev-2010}.  The main idea is to approximate the Slater density function by a sum of Gaussians
\begin{equation}
\label{gaussianExpand}
\rho(r) \approx \sum_{k=0}^{K} \lambda_k e^{-\eta_k r^2},
\qquad r=\sqrt{x^2 + y^2 + z^2}.
\end{equation}
Once the approximation \eqref{gaussianExpand} is computed, the
separation of variables in Cartesian coordinates comes for free
\begin{equation}
\label{rhoSeparated}
\rho(x,y,z) \approx \sum_{k=0}^{K} \lambda_k \,
e^{-\eta_k x^2} \, e^{-\eta_k y^2} e^{-\eta_k z^2}.
\end{equation}
The technique for the computation of the 
nodes $\lambda_k$ and the weights $\eta_k$ is
based on the computation of the inverse Laplace transform.

Let us consider a function $f_{\alpha \beta}(t)$ such that its Laplace
transform is function $F_{\alpha \beta}(s)$ 
\begin{equation}
\label{FabsqS}
F_{\alpha \beta}(s)=
\int^{\infty}_{0} e^{-st}   f_{\alpha \beta}(t) \,dt,
\end{equation}
of the following form:
\begin{equation}
\label{FabsqS1}
F_{\alpha \beta}(s)\equiv \frac{\rho(\!\sqrt{s}\,)}{C}  
={\left(\!\sqrt{s}\right)}^{\alpha} e^{-2\beta \!\sqrt{s}}
%=\int^{\infty}_{0} e^{-st}   f_{\alpha \beta}(t) \,dt,
\end{equation}
where $\alpha$ and $\beta$ are parameters
of the Slater density~\eqref{rhoDef}.
The inverse Laplace transform $f_{\alpha \beta}(x)$ can be 
computed analytically for the known $F_{\alpha \beta}(s)$. 
In \ref{App:Laplace} we present explicit expressions 
for the functions $f_{\alpha \beta}(t)$ corresponding to the 
functions \eqref{FabsqS} for integer $\alpha$ and real positive $\beta$.

Once \eqref{FabsqS} is given and the function
$f_{\alpha \beta}(t)$ is known, the 
integral~\eqref{FabsqS} is approximated by a quadrature formula
\begin{equation}
\label{lr:quadexp}
\rho(r) \approx C \sum_{k=0}^K 
w_k e^{t_k} f_{\alpha \beta}(e^{t_k}) e^{-r^2 e^{t_k}},
\end{equation}
where $w_k$ and $t_k$ are quadrature weights and nodes,
respectively. 
The procedure to compute the weights and the nodes was
taken from the paper \cite{beylkin-expsumrev-2010}. 
For the reader's convenience we give the formula 
and its derivation in ~\ref{App:rhoIntegral}.

According to equation~\eqref{gaussianExpand}
\begin{equation}
\lambda_k = C \, w_k e^{t_k} f_{\alpha \beta}(e^{t_k}),
\qquad \eta_k = e^{t_k}.
\end{equation}
It appears that only several quadrature points
(at fixed $r$) are
required to achieve the accuracy of the expansion of order
$10^{-7}$. 

\subsection{Fast computation of $T(b)$.}
\label{FastTb2DSect}
The three-dimensional integral $T(b)$ defined in~\eqref{TbDef}
can be reduced to a two-dimensional integral
by means of the decomposition~\eqref{rhoSeparated}
\begin{equation}
T(b)=\sum_{k=0}^{K} \lambda_k \iint \! \Delta E(x,z - b) \,
e^{-\eta_k x^2} e^{-\eta_k y^2} e^{-\eta_k z^2}  
 dx dy dz
\end{equation}
and analytical evaluation of the one-dimensional  Gaussian integral
\begin{equation}
\int^{\infty}_{-\infty} e^{-\eta\, y^2}dy=\sqrt{\frac{\pi}{\eta}},
\end{equation}
\begin{equation}
\label{Tb2DGauss}
T(b)=\sqrt{\pi}
\sum_{k=0}^{K} \frac{\lambda_k}{\sqrt{\eta_k}} 
\iint \! \Delta E(x,z - b) \, e^{-\eta_k x^2}  e^{-\eta_k z^2} dx dz.
\end{equation}
Suppose that $\Delta E(x,z-b)$ has been decomposed as follows
\begin{equation}
\label{DeltaEDecomp}
\Delta E(x, z - b) \approx 
\sum_{\alpha=1}^r \sigma_{\alpha} u_{\alpha}(x) g_{\alpha}(z-b).
\end{equation}
Then the integration \eqref{Tb2DGauss} can be reduced to a sequence of
one-dimensional integrations. 
\begin{equation}
\label{Tb1d1dsum}
T(b)=\sqrt{\pi}
\sum_{k=0}^{K} \frac{\lambda_k}{\sqrt{\eta_k}}
\sum_{\alpha=1}^r \sigma_{\alpha} 
I_{\alpha k} J_{\alpha k}(b),
\end{equation}
\begin{equation}
\label{If1D}
I_{\alpha k}=\int_{a_x}^{b_x} \! u_{\alpha}(x) e^{-\eta_k x^2} dx,
\end{equation}
\begin{equation}
\label{Ig1D}
J_{\alpha k}(b)=
\int_{a_y}^{b_y} \! g_{\alpha}(z-b) e^{-\eta_k z^2} dz.
\end{equation}
For the numerical approximation of the integrals \eqref{If1D} and
\eqref{Ig1D} 
we use the quadrature formula with uniform quadrature nodes (although
any suitable quadrature can be used)
\begin{equation}
\label{NumIka}
I_{\alpha k}=
\sum_{i} w^{(x)}_{i} u_{\alpha}(x_i) e^{-\eta_k x_i^2},
\end{equation}
\begin{equation}
\label{xiPoints}
x_i=-a_x + i \, h_x, \qquad
0 \le i \le 2N_x, \qquad
h_x = a_x/N_x,
\end{equation}
\begin{equation}
\label{Jzmb}
J_{\alpha k}(b)=
\sum_{j} 
w^{(z)}_j \,g_{\alpha}(z_j-b) e^{-\eta_k z_j^2},
\end{equation}
\begin{equation}
\label{zjPoints}
z_j=-a_z + j \, h_z, \qquad
0 \le j \le 2N_z, \qquad
h_z = a_z/N_z.
\end{equation}
We sample the impact parameter $b$ (which can take only positive
values) with the \emph{same step} $h_z$
\begin{equation}
\label{bPoints}
b_l= l \, h_z, \qquad
0 \le l \le N_z.
\end{equation}
This allows us to introduce a new variable $\tilde z = z - b$
discretized as 
\begin{equation}
\label{ztildakPoints}
\tilde z_k=-2a_z + k \, h_z, \qquad
0 \le k \le 3N_z,
\end{equation}
and such that for the boundary conditions \eqref{zjPoints},
\eqref{bPoints}, \eqref{ztildakPoints}
\begin{equation}
z_j - b_l = \tilde z_{j-l+N_z}.
\end{equation}
The approximation problem \eqref{DeltaEDecomp} reduces to a low-rank approximation
of the extended $(2N_x + 1) \times (3N_z + 1)$ matrix
\begin{equation}
\label{Eextand}
\Delta E(x_i, \tilde z_j)\approx
\sum_{\alpha=1}^r \sigma_{\alpha} u_{\alpha}(x_i) g_{\alpha}(\tilde z_j).
\end{equation}
This should be done only once (using the cross approximation
algorithm), and the final approximation of the integral \eqref{Jzmb} reads
\begin{equation}
\label{Jakblfast}
J_{\alpha k}(b_l) \approx
\sum_{j} 
w^{(z)}_j \,g_{\alpha}(\tilde z_{j-l+N_z}) e^{-\eta_k \tilde z_j^2}.
\end{equation}
The calculation of $T(b)$ can be summarized in 
the following algorithm.
\begin{algorithmic}[1]
\FOR{every $\gamma$-shell of the projectile ion}
\STATE compute the decomposition~\eqref{gaussianExpand} for $\rho(r)$
\STATE compute the cross approximation for the matrix $\Delta E(x_i,\tilde z_j)$
defined in~\eqref{Eextand} 
\FOR {$k=0 \ldots K$} 
\FOR {$\alpha=1 \ldots r$}
\STATE compute the integral $I_{\alpha k}$ defined in~\eqref{NumIka}
\ENDFOR
\ENDFOR
\FOR{every $b_l$ required}
%\FOR{every $\gamma$-shell of the projectile ion}
\FOR {$k=0 \ldots K$} 
\FOR {$\alpha=1 \ldots r$}
\STATE compute the integral $J_{\alpha k}(b_l)$ defined in~\eqref{Jakblfast}
\ENDFOR
%\ENDFOR
\ENDFOR
\STATE compute $T_{\gamma}(b_l)$, equation~\eqref{Tb1d1dsum}
%\STATE  add $T_{\gamma}(b_l)$ to $T(b_l)$
\ENDFOR
\ENDFOR
\end{algorithmic}

\begin{table}[t]
\caption{
Timings to compute  $T(b)$ at fixed $b$ are
presented for two cases: the DEPOSIT code (old) $T_D$
and the code based on the separated 
representations~\eqref{Tb1d1dsum} $T_s$.
Collision systems are the same as in Table~\ref{Table1Ranks}.
Number of terms in the expansion~\eqref{rhoSeparated} is labeled by $N_w$.
The calculations were carried out
for accuracy $\varepsilon=10^{-7}$ and 
$[-8,8]\otimes[-16,8]$ mesh with
$4097 \times 6145$ points.
The last column shows the speedup of the program.
}
\label{Table2Times}
\centering
\begin{tabular}{cccccc}
\hline
\hline
System & $\gamma$-Shell & $N_w$ & $T_s$ ($\times10^{-3}$ sec) & $T_{D}$ (sec) & $T_{D}/T_s$  \\
\hline
$Au^{26+}+ O$ & $4df^{17}$ & $74$ & $7.94$ & $3.89$ & $490$  \\
 & $4sp^{8}$ & $69$ & $4.92$ & $3.83$ & $778$  \\
 & $3d^{10}$ & $73$ & $3.59$ & $3.88$ & $1080$  \\
 & $3sp^{8}$ & $72$ & $3.81$ & $3.82$ & $1003$  \\ 
 & $2sp^{8}$ & $107$ & $2.42$ & $3.86$ & $1592$  \\
 & $1sp^{2}$ & $209$ & $1.24$ & $3.88$ & $3120$  \\  
\hline
$U^{28+}+ Xe$ & $5sp^{4}$ & $62$ & $10.1$ & $3.94$ & $390$ \\
 & $4df^{24}$ & $70$ & $6.05$ & $3.90$ & $644$ \\
 & $4sp^{8}$ & $67$ & $5.00$ & $3.94$ & $788$ \\
 & $3d^{10}$ & $71$ & $3.88$ & $3.92$ & $1011$ \\
 & $3sp^{8}$ & $70$ & $3.52$ & $3.90$ & $1106$ \\
 & $2sp^{8}$ & $105$ & $1.99$ & $3.87$ & $1945$ \\
 & $1sp^{2}$ & $207$ & $1.04$ & $3.88$ & $3723$ \\
\hline
\hline
\end{tabular}
\end{table}

\subsection{Numerical experiments.}
\label{NumericSect}
The most important parameter in \eqref{Eextand} is the rank $r$. It
determines the complexity of the algorithm (the smaller $r$, the better).
In Table \ref{Table1Ranks} we present the
ranks (and other numerical parameters) 
calculated for the energy gain
$\Delta E(x, \tilde z)$
corresponding to different ion-atomic collisions.
As it follows from the numerical experiments, the ranks are small.
It means that the cross decomposition
allows to decrease the size of the problem
from $O(n^2)$ elements to $O(r\cdot n)$ elements
where $r \ll n$.

In Table~\ref{Table2Times} we present
the program speedup 
for every atomic shell. Details are given in the caption
of the table. In sums \eqref{NumIka} and \eqref{Jakblfast} 
the terms less then $\epsilon = 10^{-20}$ were thrown out 
for every $x_i$ and $\tilde z_j$.
It is readily seen, that the use of the technique based on the separated 
representations \eqref{Tb1d1dsum}
allows to decrease the total time to compute $T(b)$ 
by a factor of $\sim 10^3$ compared to the previous version. In
practice the computational time is 
reduced from several hours to one minute or less on the
same hardware. 

\section{Conclusions and future work}
\label{ConclSect}
We proposed a new technique for the computation of three-dimensional
integrals based on low-rank and separated representation, that
significantly reduces the computational time. The general concept can
be applied to more complicated models 
(like ion-molecular collisions with electron loss and charge-changing 
processes) that lead to multidimensional integrals. 
For the multidimensional case we plan to use the fast approximation
techniques based on the tensor train (TT) format \cite{osel-tt-2011}.
\section*{Acknowledgements}
\label{AcknowledgeSect}
This research was partially supported by RFBR grants
12-01-00546-a, 14-01-00804-a, 13-01-12061-ofi-m.

\appendix
\section{Inverse Laplace transform sources} 
\label{App:Laplace}
For integer $\alpha$ and real positive $\beta$ the inverse Laplace
transform $f_{\alpha \beta}(t)$ of $F_{\alpha \beta}(s)$
from equation~\eqref{FabsqS} may be calculated analytically
and expressed via \emph{the Kummer's confluent hypergeometric function}
$M(a,b;z)$ (\cite{abramowitz-stegun}, chapter ~13) as follows
\begin{equation}
f_{\alpha\beta}(t)=
\frac{M\left(1+\frac{\alpha}{2},\frac{1}{2},-\frac{\beta^2}{t}\right)}
{t^{1+\frac{\alpha}{2}} \Gamma\left(-\frac{\alpha}{2} \right)} 
-2\,\beta\,\frac{M\left(\frac{3+\alpha}{2},\frac{3}{2},-\frac{\beta^2}{t}\right)}
{t^{\frac{3+\alpha}{2}} \Gamma\left(-\frac{1+\alpha}{2} \right)},
\end{equation}
where
\begin{equation}
M(a,b;z)=1+\frac{a}{b}\frac{z}{1!}+
\frac{a(a+1)}{b(b+1)}\frac{z^2}{2!}+\ldots
\end{equation}
and $\Gamma(x)$ is \emph{the Gamma function}.

Below we present the most interesting $f_{\alpha \beta}(t)$ explicitly.
Due to the difference of the normalization conditions
in spherical and Cartesian coordinates
for the Slater density~\eqref{rhoDef}
\begin{equation}
\rho(r)= N_{\gamma}\frac{(2\beta)^{2\mu+1}}{\Gamma({2\mu+1})} r^{2 \mu} e^{-2\beta},
\end{equation}
the parameter $\alpha$ is related to the parameter $\mu$ as follows
\begin{equation}
\alpha = 2 \mu - 2.
\end{equation}
The number of electrons in the shell $\gamma$ is labeled as~$N_{\gamma}$.
The parameter $\mu$ is greater or equal to unity.  It is an integer or half-integer depending on
\emph{the principal quantum number} $n$ and
\emph{the orbital quantum number} $l$ of the atomic shell.
Details can be found in~\cite{slater1960,shevelko1993}.
For example, $\mu_{1s^2}=1$, $\alpha=0$;
$\mu_{2sp^8}=2$, $\alpha = 2$;
$\mu_{4d^{10}}=3.5$, $\alpha = 5$.
Finally,
\begin{equation*}
f_{0 \beta}(t)= \frac{g_0\left(t/\beta^2 \right)}{ \sqrt{\pi} \, \beta^2},
\quad g_0(t)= \frac{e^{-\frac{1}{t}}}{t^{3/2}}
\end{equation*}
\begin{equation*}
f_{1 \beta}(t)= \frac{g_1\left(t/\beta^2\right)}{ 2\!\sqrt{\pi} \,\beta^3},
\quad
g_1(t) = -\frac{e^{-\frac{1}{t}}}{t^{3/2}}\left(1-  \frac{2}{t} \right)
\end{equation*}
\begin{equation*}
f_{2 \beta}(t)=\frac{3\, g_2\left(t/ \beta^2\right)} {2\!\sqrt{\pi} \, \beta^4},
\quad
g_2(t)=-\frac{e^{-\frac{1}{t}}} {t^{5/2}} \left(1-  \frac{2}{3t} \right)
\end{equation*}
\begin{equation*}
f_{3 \beta}(t)= \frac{3 \, g_3\left(t/\beta^2\right)}{4\!\sqrt{\pi}\,\beta^5},
\quad
g_3(t)=\frac{e^{-\frac{1}{t}}}{t^{5/2}}
\left(1 -  \frac{4}{t} + \frac{4}{3t^2} \right)
\end{equation*}
\begin{equation*}
f_{4 \beta}(t)=\frac{15 \,g_4\left(t/\beta^2\right)}{4 \! \sqrt{\pi} \, \beta^6},
\quad
g_{4}(t)=\frac{e^{-\frac{1}{t}}}{t^{7/2}}
\left(1 -  \frac{4}{3t} + \frac{4}{15t^2} \right)
\end{equation*}
\begin{equation*}
f_{5 \beta}(t)=\frac{15\, g_5\left(t/\beta^2 \right)}{8 \! \sqrt{\pi} \, \beta^{7}},
\quad
g_5(t)=-\frac{e^{-\frac{1}{t}}}{t^{7/2}} \left(1-  \frac{6}{t} +\frac{4}{t^2} - \frac{8}{15t^3} \right)
\end{equation*}
\begin{equation*}
f_{6 \beta}(t)=\frac{105\, g_6\left(t/\beta^2\right)}{8\! \sqrt{\pi}\, \beta^8},
\quad
g_6(t) = -\frac{e^{-\frac{1}{t}}}{ t^{9/2}}
\left(1- \frac{2}{t} +\frac{4}{5t^2}- \frac{8}{105t^3}  \right)
\end{equation*}

\section{Quadrature formula for the Laplace integral} 
\label{App:rhoIntegral}
To obtain the decomposition~\eqref{gaussianExpand}
for given $\alpha$ and $\beta$
we make a substitution $s \rightarrow s^2$ into the equation~\eqref{FabsqS1}
\begin{equation}
F_{\alpha \beta}(s^2)=s^{\alpha} e^{-2\beta s}=
\int^{\infty}_{0} e^{-s^2 x}   f_{\alpha \beta}(x) \,dx,
\end{equation}
then introduce another variable  $x=e^{t}$ 
\begin{equation}
\label{IntegralExpt}
F_{\alpha \beta}(s^2)=s^{\alpha} e^{-2\beta s}=
\int^{\infty}_{-\infty} e^{-s^2 e^{t}+t}   f_{\alpha \beta}(e^{t}) dt.
\end{equation}
Good news is that the function under the integral \eqref{IntegralExpt}
has exponential decay both in the spatial and frequency domains,
therefore the truncated trapezoidal (or more advanced) rule 
gives the optimal convergence
rate. The final approximation has the form
\begin{equation}
\label{IntegralGaussWeights}
 F_{\alpha \beta}(s^2) \approx \sum_{k=0}^K 
w_k e^{t_k} f_{\alpha \beta}(e^{t_k}) e^{-s^2 e^{t_k}},
\end{equation}
where parameters of the formula
\begin{equation}
t_k = a_t + k h_t, \quad
h_t = (b_t - a_t)/K
\end{equation}
have to be selected in such a way that the resulting quadrature
formula approximates the integral for a wide range of parameter~$s$. 
Typically, the choice $a_t \gtrsim -3$, $b_t \lesssim 45$, and $K\sim 250$
gives good accuracy ($\le 10^{-7}$).
As an example, in Table~\ref{Table2Times} the required number of terms
in sum~\eqref{IntegralGaussWeights} is presented.
Accurate error analysis can be found in \cite{beylkin-expsumrev-2010}.

%\section*{References}

\bibliographystyle{cpc}
\bibliography{bibtex/our,bibtex/tensor,refs,misha} 

\end{document}